# Persian Vowel recognition with MFCC and ANN on PCVC speech dataset

Saber Malekzadeh
Department of Computer Sciences, Faculty of Mathematics and Computer Sciences
Vali-e-Asr University of Rafsanjan
Rafsanjan, Iran
Saber.Malekzade@Gmail.com

Mohammad Hossein Gholizadeh
Department of Electrical Engineering, Faculty of Engineering
Vali-e-Asr University of Rafsanjan
Rafsanjan, Iran
gholizadeh@vru.ac.ir

Seyed Naser Razavi
Department of Computer Engineering, Faculty of Computer & Electrical Engineering
University of Tabriz
Tabriz, Iran
n.razavi@tabrizu.ac.ir

*Abstract*— **In this paper a new method for recognition of consonant-vowel phonemes combination on a new Persian speech dataset titled as PCVC (Persian Consonant-Vowel Combination) is proposed which is used to recognize Persian phonemes.**
**In PCVC dataset, there are 20 sets of audio samples from 10 speakers which are combinations of 23 consonant and 6 vowel phonemes of Persian language. In each sample, there is a combination of one vowel and one consonant. First, the consonant phoneme is pronounced and just after it, the vowel phoneme is pronounced. Each sound sample is a frame of 2 seconds of audio. In every 2 seconds, there is an average of 0.5 second speech and the rest is silence. In this paper, the proposed method is the implementations of the MFCC (Mel Frequency Cepstrum Coefficients) on every partitioned sound sample. Then, every train sample of MFCC vector is given to a multilayer perceptron feed-forward ANN (Artificial Neural Network) for training process. At the end, the test samples are examined on ANN model for phoneme recognition.**
**After training and testing process, the results are presented in recognition of vowels. Then, the average percent of recognition for vowel phonemes are computed.**

I. INTRODUCTION

In the Artificial Intelligence age, smart machines have become a part of our modern life and this has encouraged the expectations of friendly interaction with them. The speech, as a communication way, has seen the successful development of quite a number of applications using automatic speech recognition (ASR), including command and control, dictation, dialog systems for people with impairments, translation, etc. But, the actual challenge is to create inputs and use of speech to control applications and access information. Research on ASR is still an active topic to use speech as an input [1].

ASR – the recognition of the information of a speech signal and its transcription of set of characters – is the object of research for more than five decades, achieving notable results. It is only to be expected that advances in speech recognition make speech in any language, as the best language input method when the recognizers reach error rates under 1%. While digit recognition has already reached a rate of 99.6% [2], the phoneme recognition has not gone far more than 80% [3].

In any large vocabulary ASR (LVASR) systems, more than language model, the performance depends on phoneme recognizer. This is why research groups still working on developing the better phoneme recognizer systems. The phoneme recognition is, in fact, a recurrent problem for speech recognition community.

Phoneme recognition can be found in many application nowadays. In addition to some typical LVASR systems [4], it can be found in applications related to language and speaker recognition, music identification and also translation.

The challenge of building good acoustic models starts with applying good training algorithms to a suitable set of data. The dataset contains sound units which can be trained on training algorithms and is dependent to the detail of annotation of those units.

Most of datasets are not labeled at the phoneme level [5]. Thus, the PCVC dataset is employed because it is labeled at phoneme level. Also, unlike other phoneme based datasets [6], PCVC contains just 2 phoneme in every sample which makes the training and recognition process better. Since Persian vowels have obvious difference with all of consonants and also silence noise, it facilitates the process of vowels extraction. For extraction of consonants, as they are just pronounced before vowels, it is possible to separate them approximately. However, in this paper, to show the usability of phoneme recognition, especially vowel recognition on PCVC, the recognition algorithms are examined on PCVC vowels which have more samples than consonants.

This article is organized as follows. Section 2 discusses about the PCVC speech dataset which is used and presented for the first time in this paper. Section 3 proposes the preprocessing level with sound signal processing algorithms like MFCC and vowel extraction from PCVC speech dataset. Section 4 describes the artificial neural network which is applied for samples classification. Section 5 brings the conclusion and the last section is acknowledgement.

## II. PCVC DATASET

This dataset contains of 23 Persian consonants and 6 vowels which is listed in table 1. Table 1 contains the vowels and consonants just like the dataset. There are 10 speakers including 5 males and 5 females, and the sound samples are all possible combinations of vowels and consonants (138 samples for each speaker). The sample rate of all 2 seconds speech samples is 200000 which means there are 200000 audio samples in every 1 second. The "GH" ("غ") consonant is also one of the 23 consonant which is not using by some Persian speakers, but not all of them. Every sound sample is 2 seconds, which in average, 0.5 second of each sample is speech and the rest is silence.

For testing process, 2 other speakers are selected, and the test samples are just like training samples with different speakers.

**Table 1-** Phoneme List in PCVC dataset

| Persian form | English form | Persian Example |
|---|---|---|
| آ | A | آل |
| ای | I | ایل |
| او | ʊ | او |
| اَ | æ | اول |
| اِ | e | ارد |
| اُ | o | اردو |
| پ | P | پا |
| ب | B | با |
| ت | T | تا |
| د | D | دارو |
| چ | tʃ | چاقو |
| ج | dʒ | جارو |
| ک | K | کاری |
| گاری | G | گ |
| فاطمه | F | ف |
| واهمه | V | و |
| خاطره | Kh | خ |
| غاز | Gh | غ |
| ساز | S | س |
| زار | Z | ز |
| شار | ʃ | ش |
| ژاکت | ʒ | ژ |
| ماکت | M | م |
| نادی | N | ن |
| هادی | H | ه |
| لابه | L | ل |
| راهبه | R | ر |
| قاری | Q | ق |
| یاری | j | ی |

III. SPEECH SIGNAL PREPROCESSING

*A. Vowel extraction*

Every sound sample is an audio wave of 2 seconds of speech which starts with silence for at least 0.25s. Then, the consonants and the vowels are pronounced sequently. Since the 0.25s of every sound sample is silence, the maximum value of the silence intensity can be measured. This measurement can be employed to detect the vowels which have higher intensity than both the silence and consonants. Thus, the vowels are parts of the speech which their intensity is more than twice the noise intensity. The value of double is a suitable benchmark for detection of vowels from other elements on PCVC dataset. This part of speech is enough to detect the vowels in sound signal, but for more accuracy, it is better to use some milliseconds of the sound just before and after of this part with respect to experience for any of the vowels.

*B. Mel frequency cepstrum coefficients*

In this paper, MFCC algorithm is used to extraction of features from sound samples. MFCC is used as one of the best sound feature extraction algorithms for decades [7].

MFCC actually is a sound feature extraction algorithm which gives the ability of transform sound features from temporal domain to frequency domain and from frequency domain to time-frequency domain [8]. On time-frequency diagram, some features like shape of formants, distance and also Curvature of formants can be found that are the vital features for vowel recognition. The mentioned features are the results of human local folds which are always creating formants in different shapes when vowels are being pronounced [9, 10].

MFCCs are commonly derived as follows:

1. Take the Fourier transform of (a windowed excerpt of) a signal.

2. Map the powers of the spectrum obtained above onto the mel scale, using triangular overlapping windows.

3. Take the logarithms of the powers at each of the mel frequencies.

4. Take the discrete cosine transform of the list of mel log powers, as if it were a signal.

5. The MFCCs are the amplitudes of the resulting spectrum. [11, 12]

In this paper, this algorithm is utilized for extraction of the spectral and cepstral features of sound samples. For this purpose, the window length of 20ms is used by 10ms time step between two adjacent windows. Each window includes 50 cepstrum coefficients and 100 bands. These parameters are identified as a suitable choice in tests.

IV. ARTIFICIAL NEURAL NETWORK

A feedforward neural network is a kind of artificial neural networks where connections between the layers don't form a cycle. In a feed forward network all information always moves one forward. MLP networks consists of multiple layers, usually connected in a feed-forward way. Each neuron in one layer has connections to the neurons of the next layer. In many applications of feedforward networks, units apply a sigmoid function as an activation function. [13]

*C. Training process*

To train vowels sound features, MLP artificial neural network is used. The network is a feed forward network with Scaled conjugate gradient backpropagation training function and "MSE" performance function. It contains of 3 layers with one hidden layer, includes 50 neurons. The learning rate of network is 0.1 and regularization ratio is set to 0.5, which gives equal weight to the mean square errors and the mean square weights.

*D. Testing process*

To implement the testing process, firstly, all processes described in 3.A are applied on the test samples. Then, each sound sample pronounced by a test speaker with all possible combinations of vowels and consonants is examined and the pronounced vowel is predicted by simulation. The results are presented in table 2.

**Table 2-** Recognition percent of vowels

| Persian form | English form | Recognition Percent |
|---|---|---|
| آ | A | 80 |
| ای | I | 96 |
| او | ʊ | 96 |
| أ | æ | 100 |
| إ | e | 100 |
| أ | o | 92 |

## V. CONCLUSION

In this paper a new method is proposed for recognition of phonemes in Persian language on PCVC which is a new Persian phoneme dataset. As the results shown in table 2, the capability of the proposed vowels recognition system in predicting the phonemes are better than the other human vowels recognition methods which are proposed in the conventional approaches.

## VI. ACKNOWLEDGMENT

So many thanks to those helped us to develop PCVC dataset especially speakers: Farideh Jabraili, Hedayat Malekzadeh, Hamed Afjuland, Mohammad Ataeizadeh, Tahereh Salari, Alireza Aghaei, Parisa Seyfpour, Sahel Soltani, Mina Bayarash.

Also especial thanks to Prof.Beigi for their help about vowels recognition.

## VII. REFERENCES

1. Li, Jinyu. Soft margin estimation for automatic speech recognition. Diss. Georgia Institute of Technology, 2008.

2. Baker, Janet M., et al. "Developments and directions in speech recognition and understanding, Part 1 [DSP Education]." IEEE Signal Processing Magazine 26.3 2009.

3. Mohamed, Abdel-rahman, George E. Dahl, and Geoffrey Hinton. "Acoustic modeling using deep belief networks." IEEE Transactions on Audio, Speech, and Language Processing 20.1 2012: 14-22.

4. Morris, Jeremy, and Eric Fosler-Lussier. "Conditional random fields for integrating local discriminative classifiers." IEEE Transactions on Audio, Speech, and Language Processing 16.3 2008: 617-628.

5. Lopes, Carla, and Fernando Perdigao. "Phoneme recognition on the TIMIT database." Speech Technologies. InTech, 2011.

6. Graves, Alex, Abdel-rahman Mohamed, and Geoffrey Hinton. "Speech recognition with deep recurrent neural networks." Acoustics, speech and signal processing (icassp), 2013 ieee international conference on. IEEE, 2013.

7. Beigi, Homayoon. Fundamentals of speaker recognition. Springer Science & Business Media, 2011.

8. Ghoraani, Behnaz, and Sridhar Krishnan. "Time–frequency matrix feature extraction and classification of environmental audio signals." IEEE transactions on audio, speech, and language processing 19.7 2011: 2197-2209.

9. Palmer, J.M. Lippincott Williams and Wilkins. Anatomy for Speech and Hearing. 1993.

10. Williams, A. Lynn, Sharynne McLeod, and Rebecca J. McCauley. Interventions for Speech Sound Disorders in Children. Brookes Publishing Company. PO Box 10624, Baltimore, MD 21285, 2010.

11. Xu, Min, et al. "HMM-based audio keyword generation." Pacific-Rim Conference on Multimedia. Springer, Berlin, Heidelberg, 2004.

12. Sahidullah, Md, and Goutam Saha. "Design, analysis and experimental evaluation of block based transformation in MFCC computation for speaker recognition." Speech Communication 54.4 2012: 543-565.

13. Hornik, Kurt, Maxwell Stinchcombe, and Halbert White. "Multilayer feedforward networks are universal approximators." Neural networks 2.5 1989: 359-366. APA.